# Coupling spins to nanomechanical resonators: Toward quantum spin-mechanics

Hailin Wang* and Ignas Lekavicius

Department of Physics, University of Oregon, Eugene, OR 97403, USA

## Abstract

Spin-mechanics studies interactions between spin systems and mechanical vibrations in a nanomechanical resonator and explores their potential applications in quantum information processing. In this tutorial, we summarize various types of spin-mechanical resonators and discuss both the cavity-QED-like and the trapped-ion-like spin-mechanical coupling processes. The implementation of these processes using negatively charged nitrogen vacancy and silicon vacancy centers in diamond is reviewed. Prospects for reaching the full quantum regime of spin-mechanics, in which quantum control can occur at the level of both single spin and single phonon, are discussed with an emphasis on the crucial role of strain coupling to the orbital degrees of freedom of the defect centers.

*Corresponding author. Email: hailin@uoregon.edu

## 1) Introduction

A spin-mechanical resonator, in which electron spins couple to a mechanical mode with a high quality factor as illustrated in Fig. 1a, provides an experimental platform for quantum control of both spin and mechanical degrees of freedom. This system resembles the well-known cavity QED system that couples atoms, including artificial atoms, to cavity photons[1-3]. A spin-mechanical system can thus be viewed effectively as a phononic cavity QED system[4]. In such a system, the vibration of the mechanical oscillator enables quantum control of the spins[5-11]. Conversely, the mechanical vibration can in principle be controlled through its coupling to the spins[12-15]. Mechanical vibrations can also be used to mediate coherent interactions between spins[16-25].

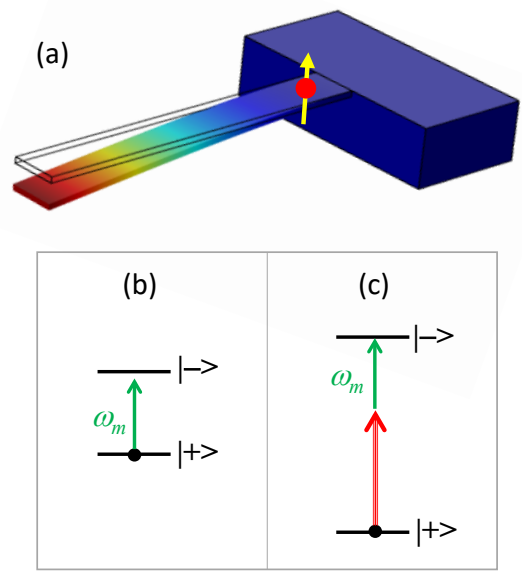

Fig. 1 (a) Schematic illustrating a spin coupling to mechanical strain of a cantilever. (b) Direct mechanically driven spin transition. (c) Sideband (i.e. phonon-assisted) spin transition, which can be driven by either microwave or optical fields.

There are, however, important differences between spin-mechanical and cavity QED systems. Compared with optical waves, mechanical waves cannot propagate in vacuum and are consequently immune to scattering losses into vacuum, a loss mechanism that is unavoidable in nano-optical systems. Nanomechanical systems can thus feature loss rates that are many orders of magnitude smaller than those of their optical counterparts, though quality factors of nanomechanical and nano-optical resonators can be comparable. In addition, mechanical vibrations can couple to nearly any type of quantum systems, including charge, spin, superconducting as well as optical qubits[26-32]. These special properties of the mechanical degrees of freedom make them highly desirable for applications such as buses or transducers in quantum networks.

When a spin couples to a mechanical oscillator, the mechanical strain can induce state mixings as well as energy shifts of the spin states, as discussed in a comprehensive earlier review on spin-mechanics[33]. The strain-induced state-mixings can lead to a mechanically driven transition between the relevant spin states (see Fig. 1b). This transition can be described by the Jaynes-Cummings Hamiltonian, which is widely used in cavity QED. In comparison, the strain induced energy shifts can lead to sideband (i.e. phonon-assisted) transitions driven by either optical or microwave fields (see Fig. 1c). The sideband transitions are analogous to those in trapped ion systems[34, 35]. A spin-mechanical resonator featuring sideband transitions can thus be viewed as a solid-state analog of trapped ions[9, 12, 36]. Sideband transitions can provide greater flexibility than direct mechanically driven transitions for the quantum control of mechanical as well as spin degrees of freedom. The sideband transitions, however, are considerably weaker than the relevant direct transitions. A spin-mechanical resonator can be configured to behave like either a cavity QED or a trapped ion system.

The experimental pursuit of spin-mechanics has been in part stimulated by the remarkable recent advances in two closely related fields, cavity optomechanics and circuit-QED based quantum acoustics. For example, quantum control of mechanical systems at the single-phonon level, including the generation of phononic Fock states, has been demonstrated by coupling a mechanical oscillator to a superconducting qubit or to a photonic crystal optical resonator[37-40]. Phonon-mediated entanglement between two superconducting qubits as well as entanglement between two distant mechanical oscillators has also been realized[41, 42].

A wide variety of solid-state spin systems can be incorporated in spin-mechanical resonators. Defect spins in silica as well as epitaxially grown semiconductor quantum dots (QDs) have been used in earlier studies[43, 44]. Color centers in diamond, which have emerged as promising qubit platforms for quantum information processing[45-48], have recently attracted strong interest for spin-mechanics studies. Negatively charged nitrogen vacancy (NV) centers feature long spin decoherence times for electron spins even at room temperature. In comparison, interstitial group IV color centers in diamond, such as negatively charged silicon vacancy (SiV), germanium vacancy, tin vacancy, and lead vacancy centers, feature relatively strong strain coupling to ground-state orbital degrees of freedom and as a result relatively short spin decoherence times at room temperature[48]. Though their room temperature spin properties are inferior to the NV center, the group IV color centers exhibit superior optical properties, with a dominant zero-phonon line in their fluorescence spectra. In addition, defect centers in semiconductors such as silicon and silicon carbide, which allow large-scale nanofabrication, can also serve as excellent spin systems for spin-mechanics studies[47].

In the following, we first summarize different types of nanomechanical resonators used for spin-mechanics studies. We then discuss spin-mechanical coupling processes including direct mechanically driven spin transitions and sideband spin transitions and review the respective experimental implementations using color centers in diamond. A particular emphasis is on the prospects of reaching the full quantum regime of spin-mechanics, in which quantum control can occur at the level of both single spin and single phonon, and on the crucial role of orbital strain coupling in reaching this regime.

## 2) Nanomechanical resonators for spin-mechanics

For spin-mechanics studies, the important mechanical parameters of a nanomechanical resonator are the mechanical frequency, $\omega_m$, quality factor, $Q_m$, and mass, $m_{eff}$. Figure 2 shows several types of nanomechanical resonators used in spin-mechanics studies. Two of the simplest and most widely used nanomechanical resonators are cantilevers and double-clamped beams (see Fig. 2a)[49-59]. The out-of-plane mechanical modes with relatively low frequencies (e.g., $\omega_m/2\pi$ < 10 MHz) in a cantilever or nanobeam can feature $Q_m$ in the order of $10^6$. Clamping losses, however, severely limit $Q_m$ of compressional in-plane mechanical modes, which can have frequencies that are orders of magnitude greater than those of the out-of-plane modes. Cantilevers and double-clamped beams are relatively easy to fabricate and characterize and are a convenient choice for spin-mechanics studies with relatively small $\omega_m$. Note that clamping losses can be suppressed via phononic engineering, for example, by embedding the nanomechanical resonator in a phononic crystal [60-62].

Structures such as microdisks feature mechanical breathing modes as well as acoustic whispering gallery modes (WGMs). Diamond microdisks with a diameter near 5 μm have been successfully fabricated (see Fig. 2b)[63]. These microdisks feature mechanical breathing modes with $\omega_m/2\pi$ near 2 GHz and $Q_m$ near $10^4$, limited by the clamping loss of the pedestal that supports the microdisk. This type of resonators also supports optical WGMs, which can couple to the mechanical breathing modes via radiation pressure force[63-66].

Optomechanical crystals, which were initially developed for cavity optomechanics studies[67], can also serve as spin-mechanical resonators. Diamond optomechanical crystals (see Fig. 2c) can feature $\omega_m/2\pi$ of a few GHz and can host NV centers as well as group-IV color centers [68, 69]. The difficulty of diamond nanofabrication has limited $Q_m$ to below $10^4$, which is considerably smaller than that has been achieved for silicon optomechanical crystals[61]. For both optomechanical crystals and microdisks, optomechanical processes can be used for the excitation and characterization of mechanical modes and can also be combined with spin-mechanical processes for potential applications such as coupling to an optical quantum network[66].

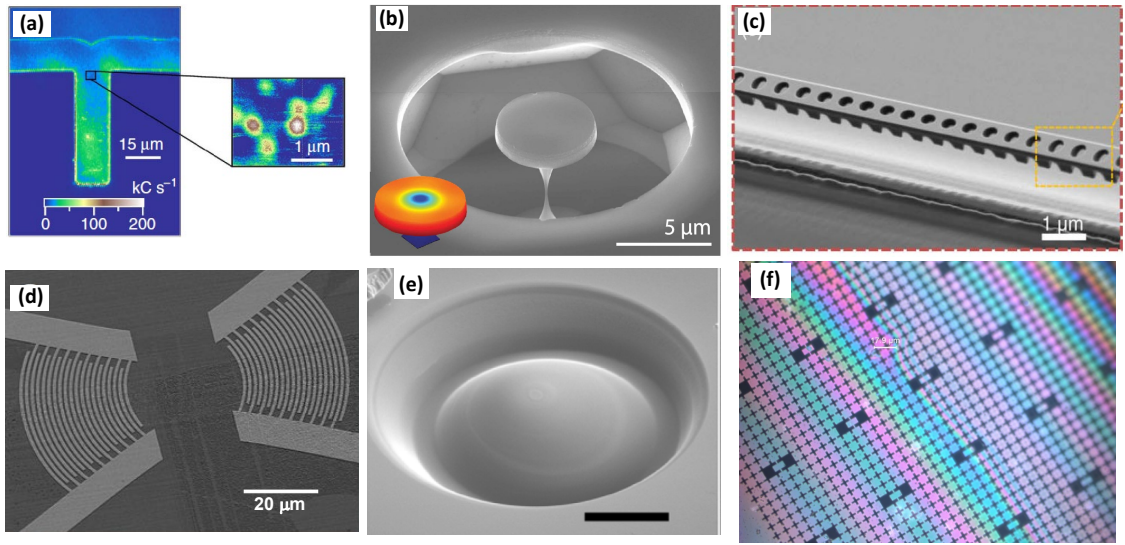

Fig. 2 (a) Confocal optical image of a diamond cantilever with NV centers. Adapted with permission from [53], licensed under a Creative Commons Attribution (CC BY) license. b) Diamond microdisk resonator. Adapted with permission from [63]. Copyright (2016) Optical Society of America. (c) Diamond optomechanical crystal. Adapted with permission from [68]. Copyright (2016) Optical Society of America. (d) SAW resonator with curved ITDs. Adapted with permission from [11], licensed under a Creative Commons Attribution (CC BY) license. (e) Diamond bulk acoustic resonator with a radius of curvature of 20 μm for the hemispherical top surface. Adapted with permission from [70]. Copyright (2019) American Chemical Society. (f) Optical image of diamond Lamb wave resonators (9.5 μm by 4.5 μm) embedded in a phononic crystal square lattice with a period of 8 μm and with a membrane thickness near 1 μm. The color fringes are due to slight variations in the membrane thickness. Adapted from [71], with the permission of AIP Publishing.

Considerable successes of spin-mechanics studies and especially circuit-QED based quantum acoustics studies have been achieved with surface acoustic wave (SAW) resonators[10, 11, 39, 72, 73]. In these types of resonators (see Fig. 2d), SAWs, which propagate on the surface of an elastic materials, are confined between two distributed Bragg reflectors, which are typically made with metallic interdigital transducers (IDTs) patterned on a piezoelectric material. Curved IDTs have been developed for Gaussian confinement of the mechanical modes [10]. Diamond and SiC based SAW resonators, with $\omega_m/2\pi$ ranging from a few hundred MHz to a few GHz and with $Q_m$ as high as a few times of $10^4$, have been developed for spin-mechanics studies. Much higher $Q_m$ has been achieved for SAW resonators used in circuit-QED based quantum acoustics studies[39, 74].

Bulk acoustic resonators (BARs), which resemble optical Fabry-Perot cavities, are natural mechanical resonators. In these resonators, the relevant mechanical vibrations are confined between two surfaces. Strong 3D spatial confinement of the mechanical modes can be achieved when one of the surfaces is suitably curved (see Fig. 3e)[70]. Materials-loss limited $Q_m$ can in principle be achieved with BAR resonators[75]. BARs, however, feature a relatively large size and correspondingly a relatively large $m_{eff}$.

Similar to BARs, a thin elastic plate with free boundaries can also serve as a natural mechanical resonator. Compressional in-plane modes in these Lamb wave resonators have relatively high $\omega_m$. For example, a thin, rectangular diamond plate with a length of 9 μm features a fundamental symmetric compressional mode with $\omega_m/2\pi$ near 1 GHz. Mechanical loss due to tethers that support the thin plate can limit $Q_m$ of a Lamb wave resonator. This anchor loss, however, can be greatly reduced by embedding the resonator in an artificial crystal lattice with a phononic band gap that spans the frequencies of the relevant Lamb modes (see Fig. 2f)[71]. Lamb wave resonators have been widely used in micro-electromechanical systems, for which the excitation and characterization of the mechanical modes are carried out with the use of IDTs. New excitation and characterization techniques are needed when it is difficult or detrimental to incorporate IDTs into the resonator structure.

In addition to the monolithic spin-mechanical systems shown in Fig. 2, composite spin-mechanical systems have also been extensively used. For example, spins near the surface of a bulk material can be coupled to a sharp ferromagnetic tip mounted on a mechanical resonator, such as a cantilever[5, 43, 76]. In addition, self-assembled QD nanowires and optically levitated diamond nanoparticles have also been explored as spin-mechanical systems[77, 78].

The two primary loss mechanisms for a mechanical resonator are the clamping or anchor loss and the intrinsic material loss due to acoustic attenuation. For many nanomechanical systems, such as nanobeams, optomechanical crystals, and Lamb wave resonators, the clamping loss can be greatly reduced via phononic band gap engineering, i.e., embedding the resonators in a phononic crystal, as discussed earlier. For materials such as silicon and diamond, the intrinsic material loss can become insignificant at low temperature.

### 3) Spin-mechanics with direct mechanically driven transitions

#### i) Analogy to cavity QED

For simplicity, we consider a two-level spin system, described by $|+\rangle$ and $|-\rangle$, coupling to a mechanical mode of a spin-mechanical resonator through a pure strain-induced state-mixing process. The spin-mechanical interaction Hamiltonian can thus be written as

$$V_\perp = \hbar g(\hat{b}+\hat{b}^+)(|+\rangle\langle-|+|-\rangle\langle+|), \qquad (1)$$

where $\hat{b}$ is the annihilation operator for the mechanical mode and $g$ is the single-phonon spin-mechanical coupling rate. This interaction Hamiltonian can lead to the resonant transition between the two spin states. Under the rotating wave approximation (RWA), the Hamiltonian reduces to the well-known Jaynes-Cummings Hamiltonian. The spin-mechanical coupling rate is related to the relevant strain coupling constant, $d$, by

$$g = x_{zpf} k_m d = \sqrt{\frac{\hbar}{2m_{eff}\omega_m}} k_m d \qquad (2)$$

where $k_m$ and $x_{zpf}$ are the wave number and zero-point fluctuation for the mechanical mode, respectively, and $x_{zpf} k_m$ is effectively the strain induced by the zero-point fluctuation. In addition to $g$, the dynamics of the spin-mechanical system are also characterized by $\gamma_s$, and $\gamma_m$, the full linewidth for the spin and mechanical system, respectively.

Similar to cavity QED, we can use the cooperativity, a dimensionless parameter defined as $C = 4g^2/\gamma_s\gamma_m$, to characterize the strength of the spin-phonon interaction. In the regime of $\gamma_m >$ ($g$, $\gamma_s$), the spin-phonon coupling can lead to enhanced decay of the spin system, with the effective spin linewidth given by $(1+C)\gamma_s$. This process is similar to the enhanced spontaneous emission or Purcell process in cavity QED. In the regime of $\gamma_s >$ ($g$, $\gamma_m$), the coupling of the mechanical oscillator to a single spin can lead to an increase in the damping of the mechanical system, with the effective mechanical linewidth given by $(1+C)\gamma_m$. Quantum control at the level of both single phonon and single spin can occur when $C>1$, which can be considered as the full quantum regime of spin-mechanics. Note that this regime, which requires $2g > \sqrt{\gamma_s\gamma_m}$, is more forgiving than the strong-coupling regime, which requires $2g >$ ($\gamma_s$, $\gamma_m$). The strong coupling regime is required for the formation of the Jaynes-Cummings ladder, but is not needed for certain quantum spin-mechanical processes.

## ii) Experimental realization of direct mechanically driven transitions

Although the model interaction Hamiltonian given in Eq. 1 is exceedingly simple, spin-mechanical coupling in actual experimental systems depends on the properties of the relevant spin states as well as the symmetry of the crystal. For example, NV ground states are characterized by a spin triplet with $m_s$= -1, 0, 1, as shown in Fig. 3a. NV centers have $C_{3v}$ crystal symmetry, with the NV axis orientated along the [111] direction of the diamond lattice as the symmetry axis[45]. While axial strain, which is along the NV axis, preserves the crystal symmetry and thus only induces energy shifts in the spin states, transverse strain induces a state mixing between the $m_s = \pm 1$ states. In this case, the underlying mechanical vibration can directly drive the spin transition between the $m_s = \pm 1$ states, as indicated in Fig. 3a. By measuring the effects of the strain on the coherent spin dynamics in Hahn echo and dynamical decoupling experiments, earlier studies have determined the NV ground-state strain coupling constants to be $d_\parallel/2\pi$ = 13.3 GHz and $d_\perp/2\pi$ =21.5 GHz, for the axial and transverse strain, respectively[53]. Slightly different results were obtained in a related experiment [52]. Note that mechanically driven transitions between the $m_s = 0$ and $m_s = \pm 1$ states have also been theoretically predicted and experimentally observed, indicating the feasibility of all-acoustic quantum control of NV spin states [79, 80]. In addition, a detailed theoretical analysis of the NV ground-state spin-strain coupling in diamond nanomechanical systems suggests the potential for sensing applications [81].

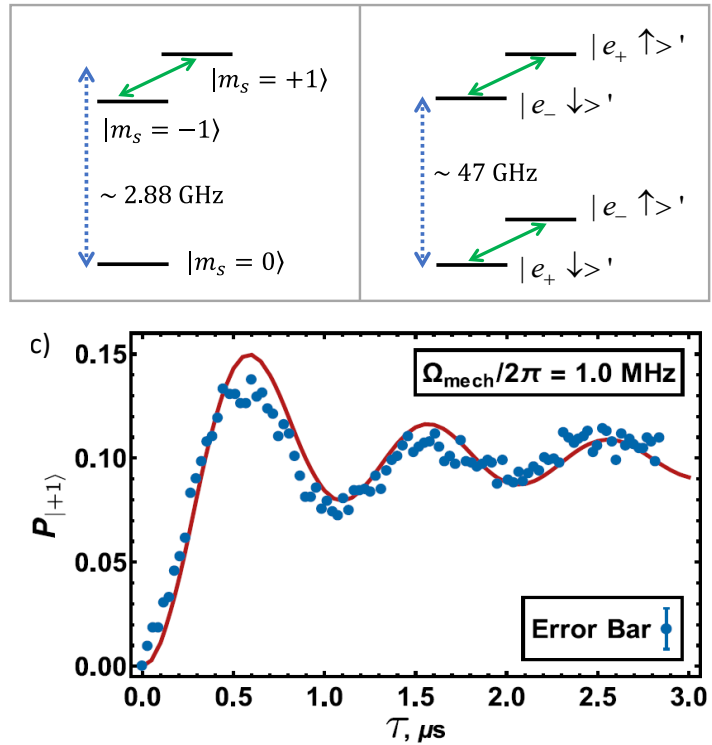

Fig. 3 (a) Schematic of the mechanically driven transition between the $m_s = \pm 1$ ground spin states in a NV center. The splitting between the two spin states is set by a magnetic field along the NV axis. (b) Mechanically driven transitions between mixed spin states in SiV ground-state doublets. The SiV center is subject to an off-axis magnetic field, which induces spin mixing as well as lifts the doublet degeneracy. (c) Mechanically driven Rabi oscillations between the $m_s = \pm 1$ states for a NV spin ensemble in a diamond BAR with $\omega_m/2\pi$=771 MHz. Adapted with the permission from [7]. Copyright (2015) Optical Society of America.

In comparison, SiV centers have $D_{3d}$ crystal symmetry, with the SiV axis as the symmetry axis. SiV ground states are characterized by orbital components, $|e_x\rangle$ and $|e_y\rangle$, as well as spin components, $|\uparrow\rangle$ and $|\downarrow\rangle$. Spin-orbit coupling leads to the formation of two doublets. In the absence of mechanical strain and external magnetic fields, the degenerate upper doublet, featuring $|e_+\rangle|\uparrow\rangle$ and $|e_-\rangle|\downarrow\rangle$, is separated by $\lambda_{so}/2\pi=47$ GHz from the degenerate lower doublet, featuring $|e_-\rangle|\uparrow\rangle$, $|e_+\rangle|\downarrow\rangle$, where $|e_\pm\rangle=(|e_x\rangle\pm i|e_y\rangle)/\sqrt{2}$ [82]. Mechanical strain can couple much more strongly to the orbital components, i.e., the spatial degrees of freedom, than to the spin components. The orbital strain coupling, however, cannot induce state-mixing between the two states in a given doublet, since these states have different spin projections. By applying an off-axis magnetic field, we can not only lift the doublet degeneracy, but also induce a spin mixing between states in different doublets. Transitions between the mixed spin states in a given doublet can now take place via the orbital strain coupling, as illustrated in Fig. 3b, where the prime symbol indicates states with spin mixing. The effective spin-mechanical coupling rate is given by $g=k_m x_{zpf} d(\gamma_{SiV}B_\perp/2\lambda_{so})$, where $\gamma_{SiV}B_\perp/2\lambda_{so}$ characterizes the amount of spin-mixing, with $B_\perp$ being the transverse magnetic field and $\gamma_{SiV}$ being the SiV spin gyromagnetic ratio[83]. In this case, $d$ is an orbital strain coupling constant and $d/2\pi$ is of order 1 PHz[83].

Direct mechanically driven spin transitions have been explored in a variety of spin-mechanical systems such as cantilevers, BARs, and SAW resonators[6-8, 10, 11, 53]. NV centers have been used in diamond cantilevers and BARs. SiV centers in diamond and divacancy centers in SiC have been used in SAW resonators. Mechanically driven Rabi oscillations have been realized in all these systems[7, 8, 10, 11]. Figure 3c shows an earlier experimental demonstration of mechanically driven Rabi oscillations between the $m_s=\pm 1$ states in a NV spin ensemble in a BAR[7]. In this experiment, the mechanical oscillations or stress waves were generated electrically with electrodes patterned on the ZnO piezoelectric film deposited on the diamond surface. The energy separation between the two spin states was set by an external magnetic field along the NV axis. In addition to the Rabi oscillations, dressed spin states due to strong mechanical driving have also been observed[8]. Similar to the microwave-dressed spin states[84, 85], the acoustic-dressed spin states have also been used for the suppression of spin dephasing induced by the nuclear spin bath in diamond[8, 86].

### iii) Toward the quantum regime with *C>1*

The regime of *C>1* has thus far not been achieved for spin-mechanical systems. The strain coupling of the NV ground spin states is extremely weak (which leads to the robust NV spin coherence even at room temperature). For a numerical estimate, we consider a diamond mechanical resonator with $m_{eff}$=10 picogram and $\omega_m/2\pi$=1 GHz. With $d_\perp/2\pi$=21.5 GHz, we have $g/2\pi$ near 10 Hz. For NV-based spin-mechanical systems, it is thus difficult to achieve $C>1$ via the ground-state strain coupling.

For interstitial group IV centers in diamond, such as SiV centers, spin-orbit coupling in the ground states can enable relatively strong ground-state orbital strain coupling, as discussed above. For a SiV center in an off-axis external magnetic field and with $\gamma_{SiV}B_\perp/2\lambda_{so} \sim 0.1$, $g$ can be nearly four orders of magnitude greater than that for a NV center in an otherwise identical nanomechanical resonator. For diamond spin-mechanical resonators, group IV centers are thus a more promising choice than NV centers in achieving $C>1$ for the cavity-QED-like spin-mechanical coupling[83].

For many semiconductors, such as Si, GaAs, and InAs, the valence band features strong spin-orbit coupling, leading to the formation of the heavy hole and light hole bands, whereas the bottom of the conduction band is *s*-like. In this regard, holes and acceptors, which are associated with the valence bands, are more promising for achieving relatively strong cavity-QED-like spin-mechanical coupling than electrons and donors, which are associated with the conduction band, as indicated in a recent experimental study[58].

The composite spin-mechanical system discussed in Section 2 exploits the interaction between a spin and a magnetic field gradient. This interaction can also lead to cavity-QED-like spin-mechanical coupling. Earlier analysis suggests that $C>1$ is achievable with the composite system[13, 76]. The composite system has been used for the detection of a single spin via magnetic resonance force microscopy[43], the mechanical control of single spins[5], the sensing of mechanical motion via coherent spin dynamics[76], and quantum transducers[16].

**4) Spin-mechanics with sideband transitions**
**i) Analogy to trapped ions**

We assume that two spin states, |+> and |–>, can couple to an excited state, |e>, through two dipole optical transitions with transition frequencies $\nu_+$ and $\nu_-$, respectively, forming a Λ-type three-level system, as shown in Fig. 4a. Two optical fields, with frequency $\omega_+$ and $\omega_-$ and Rabi

frequency $\Omega_+$ and $\Omega_-$, drive the two respective transitions, with the optical interaction Hamiltonian in the RWA given by

$$V_{Opt} = (\frac{\hbar\Omega_+}{2}e^{-i\omega_+ t}|e><+| + \frac{\hbar\Omega_-}{2}e^{-i\omega_- t}|e><-|) + h.c. \qquad (3)$$

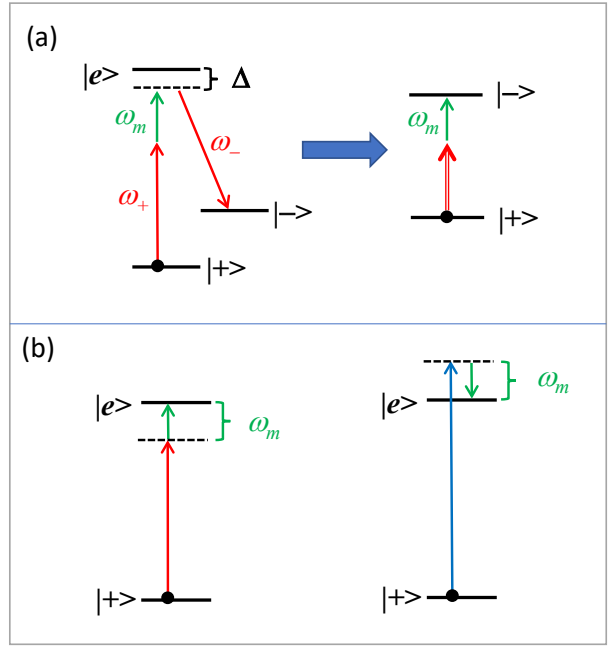

Fig. 4 (a) A $\Lambda$-type three-level system driven by two optical fields as well as a mechanical vibration leading effectively to a sideband spin transition. (b) Sideband optical transitions. Left: Red sideband. Right: Blue sideband.

For simplicity, we assume that strain coupling only leads to an energy level shift of $|e>$, with the interaction Hamiltonian given by

$$V_\| = \hbar G(\hat{b} + \hat{b}^+)|e><e|. \qquad (4)$$

where $G = k_m x_{zpf} D$, with $D$ being the deformation potential. The mechanical interaction Hamiltonian can be diagonalized with the Schrieffer-Wolff transformation, $U = \exp[\eta(\hat{b}^+ - \hat{b})|e><e|]$, where $\eta = G/\omega_m$[12, 18]. The overall effective interaction Hamiltonian after the transformation is then given by

$$V_{eff} = [\frac{\hbar\Omega_+}{2}e^{-i\omega_+ t+\eta(\hat{b}^+ -\hat{b})}|e><+|+h.c.]$$
$$+[\frac{\hbar\Omega_-}{2}e^{-i\omega_- t+\eta(\hat{b}^+ -\hat{b})}|e><-|+h.c.]-\frac{\hbar G^2}{\omega_m}|e><e| \quad (5)$$

Equation 5 has a form that is similar to the interaction Hamiltonian for trapped ions[34]. Note that $\eta=G/\omega_m$, which is generally much smaller than 1, resembles the Lamb-Dicke parameter for trapped ions. Because of the last term in Eq. 5 (a polaron shift), the effective optical transition frequencies are now $\tilde{\nu}_\pm = \nu_\pm - G^2/\omega_m$.

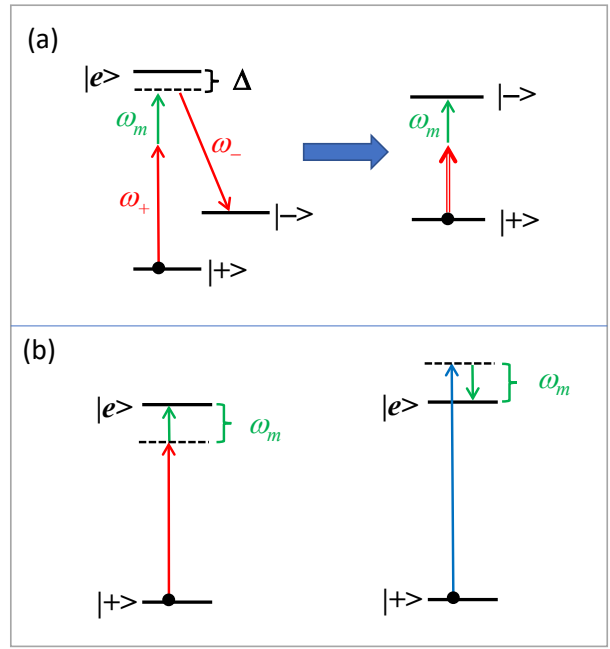

Fig. 4 (a) A Λ-type three-level system driven by two optical fields as well as a mechanical vibration leading effectively to a sideband spin transition. (b) Sideband optical transitions. Left: Red sideband. Right: Blue sideband.

In the limit that $\omega_+$ is tuned to near the red sideband of the $|+>$ to $|e>$ transition, i.e., near $\tilde{\nu}_+ - \omega_m$ (see Fig. 4b), we keep only the resonant terms in Eq. 5. In this case, the blue-sideband transitions (see Fig. 4b) are off-resonant. These transitions can be ignored if the spin-mechanical system is in the so-called resolved sideband regime, i.e., $\omega_m$ is large compared with the linewidths of the optical transitions. To the lowest order in $\eta$, the resulting effective interaction Hamiltonian can be written as,

$$V_R = [\frac{\hbar\eta\Omega_+}{2}e^{-i\omega_+ t}\hat{b}|e><+| + \frac{\hbar\Omega_-}{2}e^{-i\omega_- t}|e><-|] + h.c., \qquad (6)$$

where the first term corresponds to the first order red-sideband optical transition, with a sideband resonant condition $\omega_+ = \tilde{\nu}_+ - \omega_m$ and an effective Rabi frequency $\Omega_R = \eta\Omega_+\sqrt{n}$, with $n$ being the average phonon number. Effective Hamiltonians for higher order red sideband transitions as well as for blue sideband transitions can be similarly derived from Eq. 5.

In the limit of large dipole detuning and the optical fields being near the Raman resonance defined by $\nu_- - \omega_- = \nu_+ - \omega_+ - \omega_m$, $|e>$ can be adiabatically eliminated. In this case, Eq. 6 can be reduced to a Hamiltonian for a sideband spin transition, as illustrated schematically in Fig. 4a. The resulting interaction Hamiltonian has a form similar to that for the sideband optical transition, with a Rabi frequency $\Omega_{ss} = 2g\sqrt{n}$ where $g = \eta\Omega_+\Omega_-/(4|\Delta|)$ is the effective single-phonon spin-mechanical coupling rate, with $\Delta$ being the relevant dipole detuning[24]. The cooperativity for the spin-mechanical coupling process is still $C = 4g^2/\gamma_s\gamma_m$. Similar to trapped ion systems[34], the sideband spin as well as sideband optical transitions can be used for quantum control of both mechanical and spin degrees of freedom and for mediating coherent interactions between the spins.

An advantage of using the resonant Raman process for spin-mechanical coupling is that it is no longer necessary to match the mechanical resonance to the frequency of the spin transition since the Raman resonant condition as well as the sideband resonant condition can be achieved through the adjustment of the relevant laser frequencies. Furthermore, the spin-mechanical coupling can be turned on or off via the external laser fields. These features have been exploited in a phononic network architecture with alternating phononic crystal waveguides, in which any two adjacent mechanical resonators can form a closed mechanical subsystem[24, 25]. This architecture overcomes the inherent obstacles in scaling mechanical quantum networks and avoids the technical difficulty of employing chiral or unidirectional spin-phonon interactions.

**ii) Experimental realization of sideband transitions**

The deformation potential, *D*, is determined by the change of the band gap induced by a fractional change in the lattice constant of the crystal and can be viewed as arising from orbital strain coupling. *D* is of the order of the band gap and depends weakly on the crystal symmetry. This is in contrast to the strain-induced state mixing discussed in Section 3, which depends sensitively on the symmetry properties. The sideband optical transitions can be realized in nearly all defect centers and QDs that feature spectrally narrow zero-phonon lines. Strain-induced frequency shift of optical transitions have been characterized in detail for both NV and SiV centers[83, 87]. From these studies, we have $D/2\pi$ of order 1 PHz for both types of defect centers. The strain induced shift has also been used for the electromechanical control of defect center transition frequencies[83, 87] and the relevant decay rates of the defect centers[88].

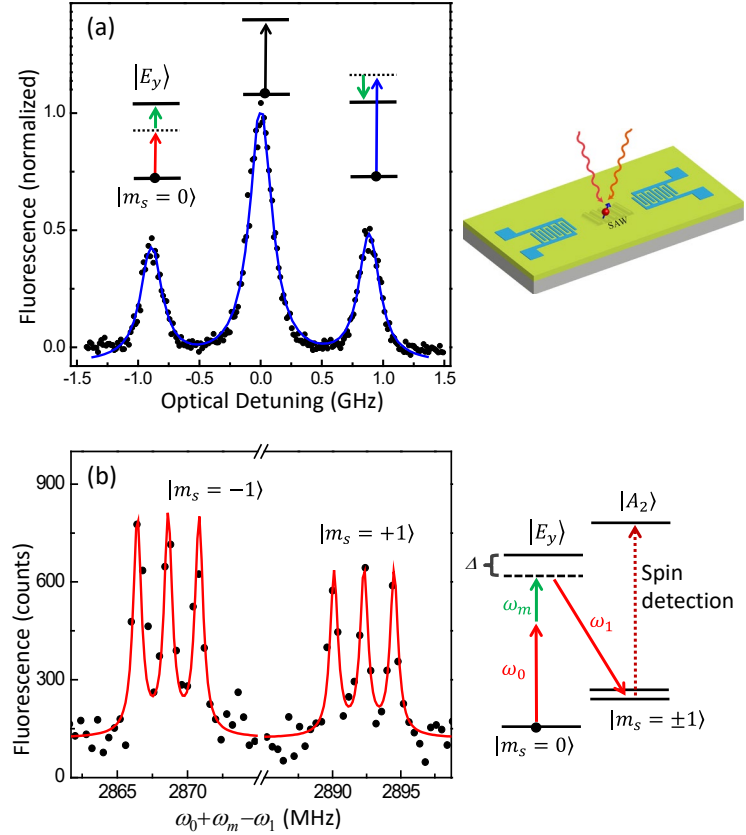

Fig. 5 (a) Optical excitation spectrum of a NV center in the presence of a SAW with $\omega_m/2\pi$=900 MHz showing both red and blue sidebands in the resolved sideband regime, as indicated in the inset. The drawing to the right illustrates a NV center coupling to both optical and SAW fields, with the SAW generated by IDTs. (b) Fluorescence from state $A_2$ of a NV center as a function of $\omega_0+\omega_m-\omega_1$ demonstrating optically driven sideband spin transitions. The energy level diagram to the right shows the transitions used for the experiment, with $\omega_m/2\pi$=818 MHz and $\Delta/2\pi$=100 MHz. Both experiments were performed at T=8 K. The figure is adapted from [9, 36].

Sideband optical transitions, including higher order transitions under relatively strong acoustic driving, have been observed in earlier studies of QDs and in both NV and SiV centers using SAW resonators as well as BARs[9, 11, 44, 89]. Figure 5a shows an experimental demonstration of sideband optical transitions for a NV center[9]. The two sidebands in the figure correspond to optical excitations via the respective red and blue sideband transitions. Rabi oscillations via sideband (i.e., phonon assisted) optical transitions have also been demonstrated for the NV center[9]. In addition, microwave-driven sideband spin transitions have been demonstrated for the $m_s = 0$ to $m_s = \pm 1$ transitions in the NV ground states via the axial strain coupling [52].

Figure 5b shows an experimental demonstration of optically driven sideband spin transitions, for which a Λ-type three-level system consisting of the $E_y$ excited state, the $m_s$=0 state, and one of the $m_s = \pm 1$ states in a NV center is used[36]. The triplet spectral feature in the figure is due to the hyperfine coupling between the electron spin and the nitrogen nuclear spin. The spectral linewidth (0.7 MHz) of the individual resonances in Fig. 5b is determined by the linewidth of the spin transition.

### iii) Toward the quantum regime with *C*>1

Optically driven sideband spin transitions can take advantage of the strong orbital strain coupling in the excited states. For a numerical estimate, we again consider a diamond mechanical resonator with $m_{eff}$=10 picogram and $\omega_m/2\pi$=1 GHz. For $\Omega_+/2\pi = \Omega_-/2\pi$=0.8 GHz and $\Delta/2\pi$=8 GHz and with $D/2\pi$= 1.1 PHz, we estimate that $g/2\pi \sim$ 10 kHz, which is promising for reaching the regime of *C*>1, if $Q_m$ of the resonator can approach $10^6$.

A limitation for optically driven spin-mechanical coupling is the decoherence induced by the optical excitation of |e>. The optical excitation rate is $\gamma_{opt} = (\Omega/2\Delta)^2 \Gamma_{ex}$, where $\Omega$ is the optical Rabi frequency and $\Gamma_{ex}$ is the excited-state population decay rate. The excited-state excitation can in principle be made negligibly small with the use of a large dipole detuning. Nevertheless, the large detuning also reduces the spin-mechanical coupling rate, which scales with $1/|\Delta|$. In addition to the use of a relatively large dipole detuning, a combination of techniques, such as adiabatic passage, including shortcuts to adiabatic passage, which employs specially-designed temporal pulse shapes to speed up the adiabatic process [90-94], can be adopted for the suppression

of optically-induced decoherence. It should be noted that the effective Lamb-Dicke parameter for a spin-mechanical system is considerably smaller than that for a trapped-ion system because of the relatively large mass of the nanomechanical resonator.

**5) Outlook**

Spin-mechanical systems with cavity-QED-like or trapped-ion-like spin-mechanical coupling are an emerging and versatile platform for exploring interactions between mechanical and spin degrees of freedom and especially for exploiting these interactions for applications in quantum science and technology. Rapid experimental advances in recent years including the exploration of a wide variety of nanomechanical resonators and spin systems have led to the experimental realization of all the important ingredients needed to take spin-mechanical systems into the full quantum regime. Given the intrinsically weak coupling between mechanical strain and pure spin degrees of freedom, future experimental systems attempting to reach this regime will likely combine relatively strong orbital strain coupling with nearly materials-loss limited nanomechanical resonators made possible by phononic band-gap engineering. The realization of the full quantum regime of spin-mechanics can open exciting opportunities, such as mechanically mediated spin entanglement and phononic quantum networks of spins[16, 18, 21, 23-25], which can serve as the building blocks for developing spin-based quantum computers. Finally, we note that a potential new avenue to further increase the spin-mechanical coupling rate at the level of zero-point quantum fluctuations is to enhance the quantum fluctuations via squeezed mechanical motion[95-97].

**Acknowledgements**

This work is supported by AFOSR and by NSF under grant No. 1719396 and No. 2012524. The data that support the findings of this study are available from the corresponding author upon reasonable request.

**References:**

1. J. M. Raimond, M. Brune, and S. Haroche, "Colloquium: Manipulating quantum entanglement with atoms and photons in a cavity," Reviews of Modern Physics **73**, 565-582 (2001).
2. H. J. Kimble, "The quantum internet," Nature **453**, 1023-1030 (2008).
3. A. Reiserer, and G. Rempe, "Cavity-based quantum networks with single atoms and optical photons," Reviews of Modern Physics **87**, 1379 (2015).
4. O. O. Soykal, R. Ruskov, and C. Tahan, "Sound-Based Analogue of Cavity Quantum Electrodynamics in Silicon," Physical Review Letters **107**, 235502 (2011).
5. S. K. Hong, M. S. Grinolds, P. Maletinsky, R. L. Walsworth, M. D. Lukin, and A. Yacoby, "Coherent, Mechanical Control of a Single Electronic Spin," Nano Letters **12**, 3920-3924 (2012).
6. E. R. MacQuarrie, T. A. Gosavi, N. R. Jungwirth, S. A. Bhave, and G. D. Fuchs, "Mechanical Spin Control of Nitrogen-Vacancy Centers in Diamond," Physical Review Letters **111**, 227602 (2013).
7. E. R. MacQuarrie, T. A. Gosavi, A. M. Moehle, N. R. Jungwirth, S. A. Bhave, and G. D. Fuchs, "Coherent control of a nitrogen-vacancy center spin ensemble with a diamond mechanical resonator," Optica **2**, 233-238 (2015).
8. A. Barfuss, J. Teissier, E. Neu, A. Nunnenkamp, and P. Maletinsky, "Strong mechanical driving of a single electron spin," Nature Physics **11**, 820 (2015).
9. D. A. Golter, T. Oo, M. Amezcua, K. A. Stewart, and H. Wang, "Optomechanical Quantum Control of a Nitrogen-Vacancy Center in Diamond," Physical Review Letters **116**, 143602 (2016).
10. S. J. Whiteley, G. Wolfowicz, C. P. Anderson, A. Bourassa, H. Ma, M. Ye, G. Koolstra, K. J. Satzinger, M. V. Holt, F. J. Heremans, A. N. Cleland, D. I. Schuster, G. Galli, and D. D. Awschalom, "Spin-phonon interactions in silicon carbide addressed by Gaussian acoustics," Nature Physics **15**, 490 (2019).
11. S. Maity, L. Shao, S. Bogdanović, S. Meesala, Y.-I. Sohn, N. Sinclair, B. Pingault, M. Chalupnik, C. Chia, L. Zheng, K. Lai, and M. Lončar, "Coherent acoustic control of a single silicon vacancy spin in diamond," Nature Communications **11**, 193 (2020).
12. I. Wilson-Rae, P. Zoller, and A. Imamoglu, "Laser cooling of a nanomechanical resonator mode to its quantum ground state," Physical Review Letters **92**, 075507 (2004).
13. P. Rabl, P. Cappellaro, M. V. G. Dutt, L. Jiang, J. R. Maze, and M. D. Lukin, "Strong magnetic coupling between an electronic spin qubit and a mechanical resonator," Physical Review B **79**, 041302 (2009).
14. K. V. Kepesidis, S. D. Bennett, S. Portolan, M. D. Lukin, and P. Rabl, "Phonon cooling and lasing with nitrogen-vacancy centers in diamond," Physical Review B **88**, 064105 (2013).
15. Z. Q. Yin, T. C. Li, X. Zhang, and L. M. Duan, "Large quantum superpositions of a levitated nanodiamond through spin-optomechanical coupling," Physical Review A **88**, 033614 (2013).
16. P. Rabl, S. J. Kolkowitz, F. H. L. Koppens, J. G. E. Harris, P. Zoller, and M. D. Lukin, "A quantum spin transducer based on nanoelectromechanical resonator arrays," Nature Physics **6**, 602-608 (2010).
17. S. J. M. Habraken, K. Stannigel, M. D. Lukin, P. Zoller, and P. Rabl, "Continuous mode cooling and phonon routers for phononic quantum networks," New Journal of Physics **14**, 115004 (2012).
18. A. Albrecht, A. Retzker, F. Jelezko, and M. B. Plenio, "Coupling of nitrogen vacancy centres in nanodiamonds by means of phonons," New Journal of Physics **15**, 083014 (2013).

19. S. D. Bennett, N. Y. Yao, J. Otterbach, P. Zoller, P. Rabl, and M. D. Lukin, "Phonon-Induced Spin-Spin Interactions in Diamond Nanostructures: Application to Spin Squeezing," Physical Review Letters **110**, 156402 (2013).
20. M. V. Gustafsson, T. Aref, A. F. Kockum, M. K. Ekstrom, G. Johansson, and P. Delsing, "Propagating phonons coupled to an artificial atom," Science **346**, 207-211 (2014).
21. M. J. A. Schuetz, E. M. Kessler, G. Giedke, L. M. K. Vandersypen, M. D. Lukin, and J. I. Cirac, "Universal Quantum Transducers Based on Surface Acoustic Waves," Physical Review X **5**, 031031 (2015).
22. Y. L. Zhang, C. L. Zou, X. B. Zou, L. Jiang, and G. C. Guo, "Phonon-induced spin squeezing based on geometric phase," Physical Review A **92**, 013825 (2015).
23. M. A. Lemonde, S. Meesala, A. Sipahigil, M. J. A. Schuetz, M. D. Lukin, M. Loncar, and P. Rabl, "Phonon Networks with Silicon-Vacancy Centers in Diamond Waveguides," Physical Review Letters **120**, 213603 (2018).
24. M. C. Kuzyk, and H. Wang, "Scaling Phononic Quantum Networks of Solid-State Spins with Closed Mechanical Subsystems," Phys. Rev. X **8**, 041027 (2018).
25. X. Li, M. C. Kuzyk, and H. Wang, "Honeycomblike Phononic Networks with Closed Mechanical Subsystems," Phys Rev Appl **11**, 064037 (2019).
26. M. Wallquist, K. Hammerer, P. Rabl, M. Lukin, and P. Zoller, "Hybrid quantum devices and quantum engineering," Physica Scripta **T137** (2009).
27. K. Stannigel, P. Rabl, A. S. Sorensen, P. Zoller, and M. D. Lukin, "Optomechanical Transducers for Long-Distance Quantum Communication," Physical Review Letters **105**, 220501 (2010).
28. L. Tian, and H. L. Wang, "Optical wavelength conversion of quantum states with optomechanics," Physical Review A **82**, 053806 (2010).
29. A. H. Safavi-Naeini, and O. Painter, "Proposal for an optomechanical traveling wave phonon-photon translator," New Journal of Physics **13**, 013017 (2011).
30. C. A. Regal, and K. W. Lehnert, "From cavity electromechanics to cavity optomechanics," Journal of Physics: Conference Series **264**, 012025 (2011).
31. Z. L. Xiang, S. Ashhab, J. Q. You, and F. Nori, "Hybrid quantum circuits: Superconducting circuits interacting with other quantum systems," Reviews of Modern Physics **85**, 623-653 (2013).
32. A. A. Clerk, K. W. Lehnert, P. Bertet, J. R. Petta, and Y. Nakamura, "Hybrid quantum systems with circuit quantum electrodynamics," Nature Physics **16**, 257-267 (2020).
33. D. Lee, K. W. Lee, J. V. Cady, P. Ovartchaiyapong, and A. C. B. Jayich, "Topical review: spins and mechanics in diamond," J Optics-Uk **19**, 033001 (2017).
34. D. Leibfried, R. Blatt, C. Monroe, and D. Wineland, "Quantum dynamics of single trapped ions," Review of Modern Physics **75**, 281-324 (2003).
35. C. Monroe, and J. Kim, "Scaling the Ion Trap Quantum Processor," Science **339**, 1164-1169 (2013).
36. D. A. Golter, T. Oo, M. Amezcua, I. Lekavicius, K. A. Stewart, and H. Wang, "Coupling a Surface Acoustic Wave to an Electron Spin in Diamond via a Dark State," Physical Review X **6**, 041060 (2016).
37. A. D. O'Connell, M. Hofheinz, M. Ansmann, R. C. Bialczak, M. Lenander, E. Lucero, M. Neeley, D. Sank, H. Wang, M. Weides, J. Wenner, J. M. Martinis, and A. N. Cleland, "Quantum ground state and single-phonon control of a mechanical resonator," Nature **464**, 697-703 (2010).

38. Y. W. Chu, P. Kharel, T. Yoon, L. Frunzio, P. T. Rakich, and R. J. Schoelkopf, "Creation and control of multi-phonon Fock states in a bulk acoustic-wave resonator," Nature **563**, 666-670 (2018).
39. K. J. Satzinger, Y. P. Zhong, H. S. Chang, G. A. Peairs, A. Bienfait, M. H. Chou, A. Y. Cleland, C. R. Conner, E. Dumur, J. Grebel, I. Gutierrez, B. H. November, R. G. Povey, S. J. Whiteley, D. D. Awschalom, D. I. Schuster, and A. N. Cleland, "Quantum control of surface acoustic-wave phonons," Nature **563**, 661-665 (2018).
40. S. Hong, R. Riedinger, I. Marinkovic, A. Wallucks, S. G. Hofer, R. A. Norte, M. Aspelmeyer, and S. Groblacher, "Hanbury Brown and Twiss interferometry of single phonons from an optomechanical resonator," Science **358**, 203-+ (2017).
41. A. Bienfait, K. J. Satzinger, Y. P. Zhong, H. S. Chang, M. H. Chou, C. R. Conner, E. Dumur, J. Grebel, G. A. Peairs, R. G. Povey, and A. N. Cleland, "Phonon-mediated quantum state transfer and remote qubit entanglement," Science **364**, 368-+ (2019).
42. R. Riedinger, A. Wallucks, I. Marinkovic, C. Loschnauer, M. Aspelmeyer, S. Hong, and S. Groblacher, "Remote quantum entanglement between two micromechanical oscillators," Nature **556**, 473-477 (2018).
43. D. Rugar, R. Budakian, H. J. Mamin, and B. W. Chui, "Single spin detection by magnetic resonance force microscopy," Nature **430**, 329-332 (2004).
44. M. Metcalfe, S. M. Carr, A. Muller, G. S. Solomon, and J. Lawall, "Resolved Sideband Emission of InAs/GaAs Quantum Dots Strained by Surface Acoustic Waves," Physical Review Letters **105**, 037401 (2010).
45. M. W. Doherty, N. B. Manson, P. Delaney, F. Jelezko, J. Wrachtrup, and L. C. L. Hollenberg, "The nitrogen-vacancy colour centre in diamond," Phys Rep **528**, 1-45 (2013).
46. L. Childress, R. Walsworth, and M. Lukin, "Atom-like crystal defects: From quantum computers to biological sensors," Phys Today **67**, 38-43 (2014).
47. D. D. Awschalom, R. Hanson, J. Wrachtrup, and B. B. Zhou, "Quantum technologies with optically interfaced solid-state spins," Nature Photonics **12**, 516-527 (2018).
48. C. Bradac, W. B. Gao, J. Forneris, M. E. Trusheim, and I. Aharonovich, "Quantum nanophotonics with group IV defects in diamond," Nature Communications **10**, 5625 (2019).
49. O. Arcizet, V. Jacques, A. Siria, P. Poncharal, P. Vincent, and S. Seidelin, "A single nitrogen-vacancy defect coupled to a nanomechanical oscillator," Nature Physics **7**, 879-883 (2011).
50. P. Ovartchaiyapong, L. M. A. Pascal, B. A. Myers, P. Lauria, and A. C. B. Jayich, "High quality factor single-crystal diamond mechanical resonator," Applied Physics Letters **101**, 163505 (2012).
51. M. J. Burek, N. P. de Leon, B. J. Shields, B. J. M. Hausmann, Y. W. Chu, Q. M. Quan, A. S. Zibrov, H. Park, M. D. Lukin, and M. Loncar, "Free-Standing Mechanical and Photonic Nanostructures in Single-Crystal Diamond," Nano Letters **12**, 6084-6089 (2012).
52. J. Teissier, A. Barfuss, P. Appel, E. Neu, and P. Maletinsky, "Strain Coupling of a Nitrogen-Vacancy Center Spin to a Diamond Mechanical Oscillator," Physical Review Letters **113**, 020503 (2014).
53. P. Ovartchaiyapong, K. W. Lee, B. A. Myers, and A. C. B. Jayich, "Dynamic strain-mediated coupling of a single diamond spin to a mechanical resonator," Nature Communications **5**, 4429 (2014).

54. Y. Tao, J. M. Boss, B. A. Moores, and C. L. Degen, "Single-crystal diamond nanomechanical resonators with quality factors exceeding one million," Nature Communications **5**, 3638 (2014).
55. I. Yeo, P. L. de Assis, A. Gloppe, E. Dupont-Ferrier, P. Verlot, N. S. Malik, E. Dupuy, J. Claudon, J. M. Gerard, A. Auffeves, G. Nogues, S. Seidelin, J. P. Poizat, O. Arcizet, and M. Richard, "Strain-mediated coupling in a quantum dot-mechanical oscillator hybrid system," Nature Nanotechnology **9**, 106-110 (2014).
56. B. Khanaliloo, H. Jayakumar, A. C. Hryciw, D. P. Lake, H. Kaviani, and P. E. Barclay, "Single-Crystal Diamond Nanobeam Waveguide Optomechanics," Physical Review X **5**, 041051 (2015).
57. S. Meesala, Y. I. Sohn, H. A. Atikian, S. Kim, M. J. Burek, J. T. Choy, and M. Loncar, "Enhanced Strain Coupling of Nitrogen-Vacancy Spins to Nanoscale Diamond Cantilevers," Phys Rev Appl **5**, 034010 (2016).
58. S. G. Carter, A. S. Bracker, G. W. Bryant, M. Kim, C. S. Kim, M. K. Zalalutdinov, M. K. Yakes, C. Czarnocki, J. Casara, M. Scheibner, and D. Gammon, "Spin-Mechanical Coupling of an InAs Quantum Dot Embedded in a Mechanical Resonator," Physical Review Letters **121**, 246801 (2018).
59. T. Oeckinghaus, S. A. Momenzadeh, P. Scheiger, T. Shalomayeya, A. Finkler, D. Dasari, R. Stohr, and J. Wrachtrup, "Spin-Phonon Interfaces in Coupled Nanomechanical Cantilevers," Nano Letters **20**, 463-469 (2020).
60. P. L. Yu, K. Cicak, N. S. Kampel, Y. Tsaturyan, T. P. Purdy, R. W. Simmonds, and C. A. Regal, "A phononic bandgap shield for high-Q membrane microresonators," Applied Physics Letters **104**, 023510 (2014).
61. S. M. Meenehan, J. D. Cohen, G. S. MacCabe, F. Marsili, M. D. Shaw, and O. Painter, "Pulsed Excitation Dynamics of an Optomechanical Crystal Resonator near Its Quantum Ground State of Motion," Phys Rev X **5**, 041002 (2015).
62. Y. Tsaturyan, A. Barg, E. S. Polzik, and A. Schliesser, "Ultracoherent nanomechanical resonators via soft clamping and dissipation dilution," Nat Nanotechnol **12**, 776 (2017).
63. M. Mitchell, B. Khanaliloo, D. P. Lake, T. Masuda, J. P. Hadden, and P. E. Barclay, "Single-crystal diamond low-dissipation cavity optomechanics," Optica **3**, 963 (2016).
64. A. Schliesser, O. Arcizet, R. Riviere, G. Anetsberger, and T. J. Kippenberg, "Resolved-sideband cooling and position measurement of a micromechanical oscillator close to the Heisenberg uncertainty limit," Nature Physics **5**, 509-514 (2009).
65. Y. S. Park, and H. L. Wang, "Resolved-sideband and cryogenic cooling of an optomechanical resonator," Nature Physics **5**, 489-493 (2009).
66. M. Aspelmeyer, T. J. Kippenberg, and F. Marquard, "Cavity optomechanics," Reviews of Modern Physics **86**, 1391-1452 (2014).
67. J. Chan, T. P. M. Alegre, A. H. Safavi-Naeini, J. T. Hill, A. Krause, S. Groblacher, M. Aspelmeyer, and O. Painter, "Laser cooling of a nanomechanical oscillator into its quantum ground state," Nature **478**, 89-92 (2011).
68. M. J. Burek, J. D. Cohen, S. M. Meenehan, N. El-Sawah, C. Chia, T. Ruelle, S. Meesala, J. Rochman, H. A. Atikian, M. Markham, D. J. Twitchen, M. D. Lukin, O. Painter, and M. Loncar, "Diamond optomechanical crystals," Optica **3**, 1404-1411 (2016).
69. J. V. Cady, O. Michel, K. W. Lee, R. N. Patel, C. J. Sarabalis, A. H. Safavi-Naeini, and A. C. B. Jayich, "Diamond optomechanical crystals with embedded nitrogen-vacancy centers," Quantum Sci Technol **4**, 024009 (2019).

70. H. Y. Chen, N. F. Opondo, B. Y. Jiang, E. R. MacQuarrie, R. S. Daveau, S. A. Bhave, and G. D. Fuchs, "Engineering Electron-Phonon Coupling of Quantum Defects to a Semiconfocal Acoustic Resonator," Nano Letters **19**, 7021-7027 (2019).
71. I. Lekavicius, T. Oo, and H. L. Wang, "Diamond Lamb wave spin-mechanical resonators with optically coherent nitrogen vacancy centers," J Appl Phys **126**, 214301 (2019).
72. F. J. R. Schulein, E. Zallo, P. Atkinson, O. G. Schmidt, R. Trotta, A. Rastelli, A. Wixforth, and H. J. Krenner, "Fourier synthesis of radiofrequency nanomechanical pulses with different shapes," Nature Nanotechnology **10**, 512-516 (2015).
73. P. Delsing, A. N. Cleland, M. J. A. Schuetz, J. Knorzer, G. Giedke, J. I. Cirac, K. Srinivasan, M. Wu, K. C. Balram, C. Bauerle, T. Meunier, C. J. B. Ford, P. V. Santos, E. Cerda-Mendez, H. L. Wang, H. J. Krenner, E. D. S. Nysten, M. Weiss, G. R. Nash, L. Thevenard, C. Gourdon, P. Rovillain, M. Marangolo, J. Y. Duquesne, G. Fischerauer, W. Ruile, A. Reiner, B. Paschke, D. Denysenko, D. Volkmer, A. Wixforth, H. Bruus, M. Wiklund, J. Reboud, J. M. Cooper, Y. Q. Fu, M. S. Brugger, F. Rehfeldt, and C. Westerhausen, "The 2019 surface acoustic waves roadmap," J Phys D Appl Phys **52**, 353001 (2019).
74. Y. W. Chu, P. Kharel, W. H. Renninger, L. D. Burkhart, L. Frunzio, P. T. Rakich, and R. J. Schoelkopf, "Quantum acoustics with superconducting qubits," Science **358**, 199-202 (2017).
75. P. Kharel, Y. W. Chu, M. Power, W. H. Renninger, R. J. Schoelkopf, and P. T. Rakich, "Ultra-high-Q phononic resonators on-chip at cryogenic temperatures," Apl Photonics **3**, 066101 (2018).
76. S. Kolkowitz, A. C. B. Jayich, Q. P. Unterreithmeier, S. D. Bennett, P. Rabl, J. G. E. Harris, and M. D. Lukin, "Coherent Sensing of a Mechanical Resonator with a Single-Spin Qubit," Science **335**, 1603-1606 (2012).
77. M. Montinaro, G. Wust, M. Munsch, Y. Fontana, E. Russo-Averchi, M. Heiss, A. F. I. Morral, R. J. Warburton, and M. Poggio, "Quantum Dot Opto-Mechanics in a Fully Self-Assembled Nanowire," Nano Letters **14**, 4454-4460 (2014).
78. T. M. Hoang, J. Ahn, J. Bang, and T. C. Li, "Electron spin control of optically levitated nanodiamonds in vacuum," Nature Communications **7** (2016).
79. P. Udvarhelyi, V. O. Shkolnikov, A. Gali, G. Burkard, and A. Palyi, "Spin-strain interaction in nitrogen-vacancy centers in diamond," Physical Review B **98**, 075201 (2018).
80. H. Y. Chen, S. A. Bhave, and G. D. Fuchs, "Acoustically Driving the Single-Quantum Spin Transition of Diamond Nitrogen-Vacancy Centers," Phys Rev Appl **13**, 054068 (2020).
81. M. S. J. Barson, P. Peddibhotla, P. Ovartchaiyapong, K. Ganesan, R. L. Taylor, M. Gebert, Z. Mielens, B. Koslowski, D. A. Simpson, L. P. McGuinness, J. McCallum, S. Prawer, S. Onoda, T. Ohshima, A. C. B. Jayich, F. Jelezko, N. B. Manson, and M. W. Doherty, "Nanomechanical Sensing Using Spins in Diamond," Nano Letters **17**, 1496-1503 (2017).
82. C. Hepp, T. Muller, V. Waselowski, J. N. Becker, B. Pingault, H. Sternschulte, D. Steinmuller-Nethl, A. Gali, J. R. Maze, M. Atature, and C. Becher, "Electronic Structure of the Silicon Vacancy Color Center in Diamond," Physical Review Letters **112**, 036405 (2014).
83. S. Meesala, Y.-I. Sohn, B. Pingault, L. Shao, H. A. Atikian, J. Holzgrafe, M. Gundogan, C. Stavrakas, A. Sipahigil, C. Chia, M. J. Burek, M. Zhang, L. Wu, J. L. Pacheco, J. Abraham, E. Bielejec, M. D. Lukin, M. Atature, and M. Loncar, "Strain engineering of the silicon-vacancy center in diamond," Physical Review B **97**, 205444 (2018).
84. X. K. Xu, Z. X. Wang, C. K. Duan, P. Huang, P. F. Wang, Y. Wang, N. Y. Xu, X. Kong, F. Z. Shi, X. Rong, and J. F. Du, "Coherence-Protected Quantum Gate by Continuous Dynamical Decoupling in Diamond," Physical Review Letters **109**, 070502 (2012).

85. D. A. Golter, T. K. Baldwin, and H. L. Wang, "Protecting a Solid-State Spin from Decoherence Using Dressed Spin States," Physical Review Letters **113**, 237601 (2014).
86. E. R. MacQuarrie, T. A. Gosavi, S. A. Bhave, and G. D. Fuchs, "Continuous dynamical decoupling of a single diamond nitrogen-vacancy center spin with a mechanical resonator," Physical Review B **92**, 224419 (2015).
87. K. W. Lee, D. Lee, P. Ovartchaiyapong, J. Minguzzi, J. R. Maze, and A. C. B. Jayich, "Strain Coupling of a Mechanical Resonator to a Single Quantum Emitter in Diamond," Phys Rev Appl **6**, 034005 (2016).
88. Y. I. Sohn, S. Meesala, B. Pingault, H. A. Atikian, J. Holzgrafe, M. Gundogan, C. Stavrakas, M. J. Stanley, A. Sipahigil, J. Choi, M. Zhang, J. L. Pacheco, J. Abraham, E. Bielejec, M. D. Lukin, M. Atature, and M. Loncar, "Controlling the coherence of a diamond spin qubit through its strain environment," Nature Communications **9**, 2012 (2018).
89. H. Y. Chen, E. R. MacQuarrie, and G. D. Fuchs, "Orbital State Manipulation of a Diamond Nitrogen-Vacancy Center Using a Mechanical Resonator," Physical Review Letters **120**, 167401 (2018).
90. X. Chen, I. Lizuain, A. Ruschhaupt, D. Guery-Odelin, and J. G. Muga, "Shortcut to Adiabatic Passage in Two- and Three-Level Atoms," Physical Review Letters **105**, 123003 (2010).
91. D. A. Golter, and H. L. Wang, "Optically Driven Rabi Oscillations and Adiabatic Passage of Single Electron Spins in Diamond," Physical Review Letters **112**, 116403 (2014).
92. B. B. Zhou, A. Baksic, H. Ribeiro, C. G. Yale, F. J. Heremans, P. C. Jerger, A. Auer, G. Burkard, A. A. Clerk, and D. D. Awschalom, "Accelerated quantum control using superadiabatic dynamics in a solid-state lambda system," Nature Physics **13**, 330 (2017).
93. H. Ribeiro, A. Baksic, and A. A. Clerk, "Systematic Magnus-Based Approach for Suppressing Leakage and Nonadiabatic Errors in Quantum Dynamics," Physical Review X **7**, 011021 (2017).
94. K. Bergmann, H. C. Nagerl, C. Panda, G. Gabrielse, E. Miloglyadov, M. Quack, G. Seyfang, G. Wichmann, S. Ospelkaus, A. Kuhn, S. Longhi, A. Szameit, P. Pirro, B. Hillebrands, X. F. Zhu, J. Zhu, M. Drewsen, W. K. Hensinger, S. Weidt, T. Halfmann, H. L. Wang, G. S. Paraoanu, N. V. Vitanov, J. Mompart, T. Busch, T. J. Barnum, D. D. Grimes, R. W. Field, M. G. Raizen, E. Narevicius, M. Auzinsh, D. Budker, A. Palffy, and C. H. Keitel, "Roadmap on STIRAP applications," J Phys B-at Mol Opt **52**, 202001 (2019).
95. H. K. Lau, and A. A. Clerk, "Ground-State Cooling and High-Fidelity Quantum Transduction via Parametrically Driven Bad-Cavity Optomechanics," Physical Review Letters **124**, 103602 (2020).
96. S. C. Burd, R. Srinivas, J. J. Bollinger, A. C. Wilson, D. J. Wineland, D. Leibfried, D. H. Slichter, and D. T. C. Allcock, "Quantum amplification of mechanical oscillator motion," Science **364**, 1163 (2019).
97. P.-B. Li, Y. Zhou, W.-B. Gao, and F. Nori, "Enhancing spin-phonon and spin-spin interactions using linear resources in a hybrid quantum system," arXiv:2003.07151 (2020).